\documentclass[runningheads]{llncs}
\usepackage[T1]{fontenc}
\usepackage{graphicx}
\begin{document}
\title{Multiscale Parallel Simulation of Malignant Pleural Mesothelioma via Adaptive Domain Partitioning – an Efficiency Analysis Study 
}
\titlerunning{Multiscale Parallel Simulation of Mesothelioma}
\author{Anton Dolganov\inst{1,2}\orcidID{0009-0001-6014-1812} \and
Valeria Krzhizhanovskaya\inst{1}\orcidID{0000-0002-8247-129X} \and
 Stefano Trebeschi\inst{2,3}\orcidID{0000-0002-5714-289X} \and
Vivek M. Sheraton\inst{1*}\orcidID{0000-0002-6577-6016} }
\authorrunning{A. Dolganov et al.}
\institute{Computational Science Lab, Informatics Institute, University of Amsterdam, The 
Netherlands
\and 
Department of Radiology, The Netherlands Cancer Institute – Antoni van 
Leeuwenhoek Hospital, Amsterdam, The Netherlands
\and
GROW – Research Institute for Oncology \& Reproduction, Maastricht University, Maastricht, the Netherlands \\
* Corresponding author: \email{v.s.muniraj@uva.nl}}
\maketitle % typeset the header of the contribution
\begin{abstract}
A novel parallel efficiency analysis on a framework for simulating the growth of Malignant Pleural Mesothelioma (MPM) tumours is presented. Proliferation of MPM tumours in the pleural space is simulated using a Cellular Potts Model (CPM) coupled with partial differential equations (PDEs). Using segmented lung data from CT scans, an environment is set up with artificial tumour data in the pleural space, representing the simulation domain, onto which a dynamic bounding box is applied to restrict computations to the region of interest, dramatically reducing memory and CPU overhead. This adaptive partitioning of the domain enables efficient use of computational resources by reducing the three-dimensional(3D) domain over which the PDEs are to be solved. The PDEs, representing oxygen, nutrients, and cytokines, are solved using the finite-volume method with a first-order implicit Euler scheme. Parallelization is realized using the public Python library \texttt{mpi4py} in combination with \texttt{LinearGMRESSolver} and \texttt{PETSc} for efficient convergence. Performance analyses have shown that parallelization achieves a reduced solving time compared to serial computation. Also, optimizations enable efficient use of the available memory and improved load balancing amongst the cores. 

\end{abstract}
\newpage
\section{Introduction}
Despite the rapid progression of technology and advancements in the medical field, cancer remains one of the most complex challenges in modern medicine. The complexity of cancer arises from both its internal heterogeneity and its dynamic interactions with the microenvironment on molecular, cellular and tissue-level, hindering a comprehensive understanding of its intricate nature. Tumour growth is an emerging phenomenon that arises from intercellular signalling dynamics between cancer cells and their microenvironments, consisting of stromal fibroblasts, immune cells, a complex vascular network and a heterogeneous environment of extracellular matrix components (ECM)~\cite{wu2014modeling,swat2012multi}. Traditional experimental approaches struggle to capture the spatiotemporal complexity of multiscale interactions. This emphasises the importance of computational models that connect molecular mechanisms to tissue-scale behaviours.

Accurate simulations and predictions of cancer progression must incorporate various phenomena on scales from nanometres to millimetres. Intracellular signalling pathways govern cell cycle regulation and adhesion molecule expression, whereas tissue-level forces and nutrient gradients shape collective migration and metastatic potential~\cite{andasari2012integrating,ma2024comprehensive}. For example, hypoxia-driven upregulation of the vascular endothelial growth factor (VEGF), which besides playing a key role in the promotion of angiogenesis also induces epithelial-mesenchymal transition (EMT), allowing individual cells to detach from the primary tumour cell clusters and invade the surrounding tissues~\cite{wu2014modeling,andasari2012integrating}. This type of intracellular to tissue transition highlights the necessity for frameworks that bridge low-level molecular signalling with high-level migration patterns. Similarly, metabolic symbiosis in tumours involves a cooperative interaction between hypoxic cancer cells that produce lactate via glycolysis, which in turn is absorbed by oxygenated cancer cells to refuel oxidative phosphorylation, which is essential for tumour cell survival and growth~\cite{jayathilake2024metabolic}. Capturing these complex interdependencies requires multiscale frameworks that can couple discrete cellular behaviours with continuum-level ODEs/PDEs, a computational challenge that demands unprecedented level and resolution~\cite{wu2014modeling,swat2012multi}.

Traditional serial computing algorithms struggle to handle the computational demands of realistic tumour models. Even 2D multiscale cancer models require extensive computational resources to simulate chemotherapy treatment, which exhausts memory resources, and incurs runtimes that span days or weeks~\cite{deisboeck2011multiscale}. The Cellular Potts Model (CPM) implemented in platforms like CompuCell3D showcases these exact challenges:
At each Monte Carlo Step (MCS), energy functions for millions of lattice sites are iteratively evaluated while tracking chemical diffusion and cell-state interactions~\cite{andasari2012integrating,swat2012multi}, whilst at tissue level, reaction-diffusion PDEs are solved for oxygen, nutrients, and growth factors, increasing the computational loads even more. All these intermediate computations of spatial gradients and cell contact histories at each MCS require the data to be stored between the steps, creating serious memory bottlenecks. 

Recent studies, which investigated spatial-temporal variations in drug efficacy within 3D tumour spheroids~\cite{kaura2021effects} or explored the effects of combined chemotherapy drugs on tumour dynamics~\cite{sheraton2020emergence}, highlight the growing complexity and associated computational intensity of modelling and simulating complicated tumour behaviour. The combination of the heavy computational load with the memory bottleneck makes the serial approach for predictive oncology impractical, especially for more complex computational problems or large 3D environments.

To reduce computational time and resolve the memory bottlenecks, parallel algorithms are commonly applied on multi-core CPUs~\cite{korkhov2008grid,krzhizhanovskaya2001distributed}, GPUs and distributed memory clusters~\cite{wu2014modeling,swat2012multi}, to decompose the computational workload across parallel threads or processes~\cite{andasari2012integrating}. For Agent Based Models (ABMs), parallelization enables simultaneous updates of cell states and (microenvironmental) variables, resulting in a drastic reduction of simulation runtime~\cite{stack2022openacc}. Particularly interesting is the distributed-memory framework that uses Message Passing Interface (MPI), which enhances the scalability in high performance computing (HPC) clusters by allowing the spatial domains to be partitioned via inter-process communication. A recent study explored such challenges of parallelizing bacterial biofilm simulations, pointing out how load imbalance and communication overhead can seriously affect simulation efficiency in these frameworks~\cite{sheraton2018parallel}. For further load balancing optimization hybrid approaches that allow the integration of MPI with shared memory can be used, in particular for heterogeneous workloads~\cite{andasari2012integrating,krzhizhanovskaya2007dynamic}, like discrete cell agents and continuum solvers as mentioned previously~\cite{macklin2009multiscale,chen2007parallel}. 

In this paper we propose an adaptive domain partitioning framework with the goal of minimizing the computational load of our 3D computational model for simulating the growth of MPM. This is achieved by (1) focusing only on the regions of interest, namely the region where the MPM is located, in our CPM with continuum PDE solvers; (2) using parallel computing and (3) applying an iterative GMRES solver. We perform a detailed parallel performance analysis with a focus on speed-up, load imbalance and time taken to solve PDEs in combination with the dynamic bounding box strategy. We demonstrate that adaptive partitioning in combination with robust parallelization can significantly enhance simulation efficiency in complex MPM models. 

\section{Methodology}
\subsection{Multiscale Model and Computational Domains}
\begin{figure}[!h]
\includegraphics[width=\textwidth]{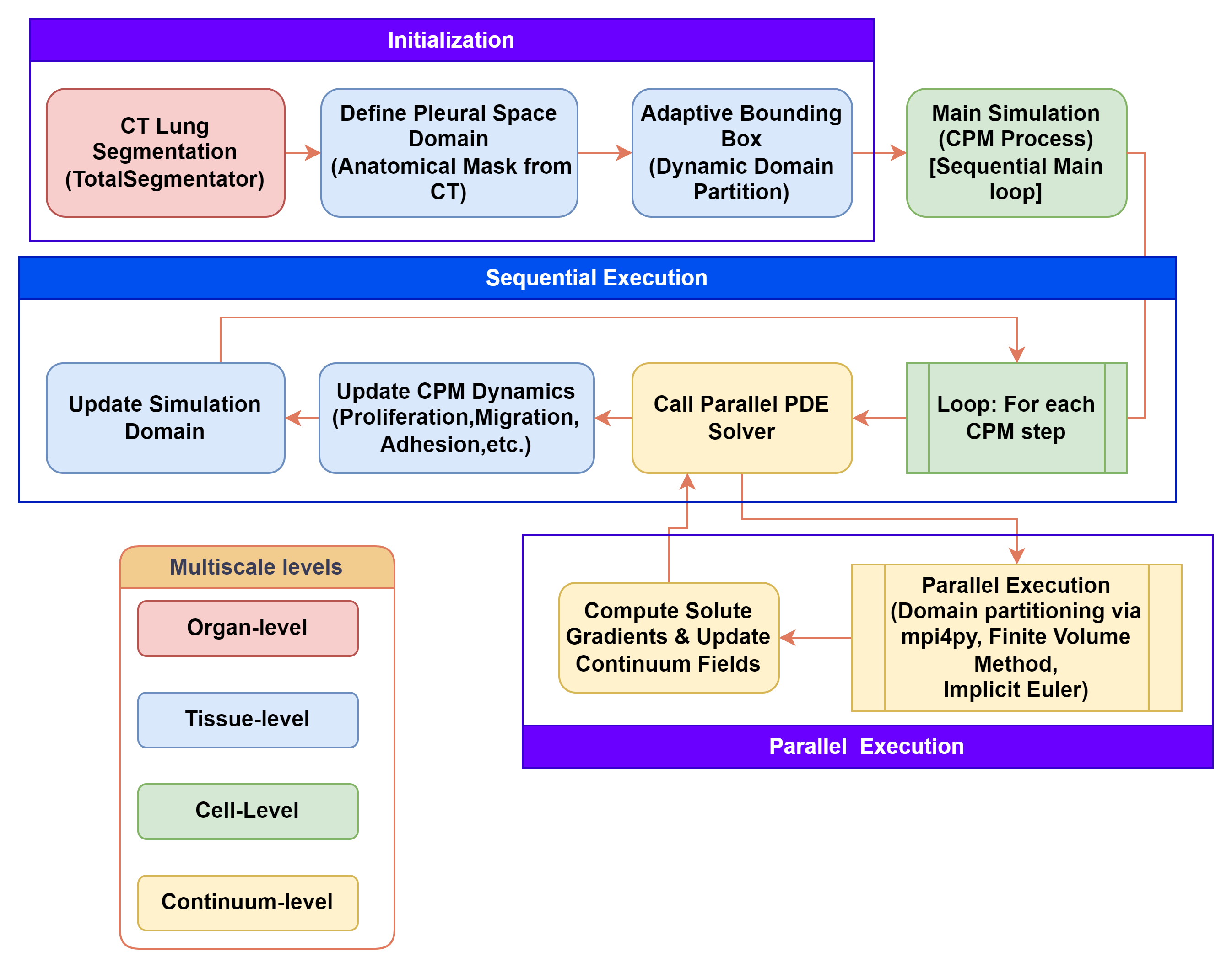}
\caption{Flowchart depicting the multiscale modelling workflow, starting from organ-level CT segmentation and progressing through tissue-level domain definition, cell-level CPM dynamics and parallel continuum-level PDE computations. Everything takes place in the Compucell3D Steppables with exception of the Parallel execution.} \label{fig1}
\end{figure}

Our framework integrates cell-level interactions with tissue-scale dynamics. This multiscale modelling covers a lower-scale process occurring over seconds, namely solute diffusion and reaction processes that shape the nutrient and chemical gradient, while also modelling cellular proliferation that unfolds over hours, resulting in growth and motility over days, linking molecular processes to tumour progression. Building on these capabilities, the goal of this study is to simulate the proliferation of MPM within the pleural space, the region between the parietal and visceral pleura of the lungs. This multiscale workflow is depicted in Fig.~\ref{fig1}.

\begin{figure}[!h]
\includegraphics[width=\textwidth]{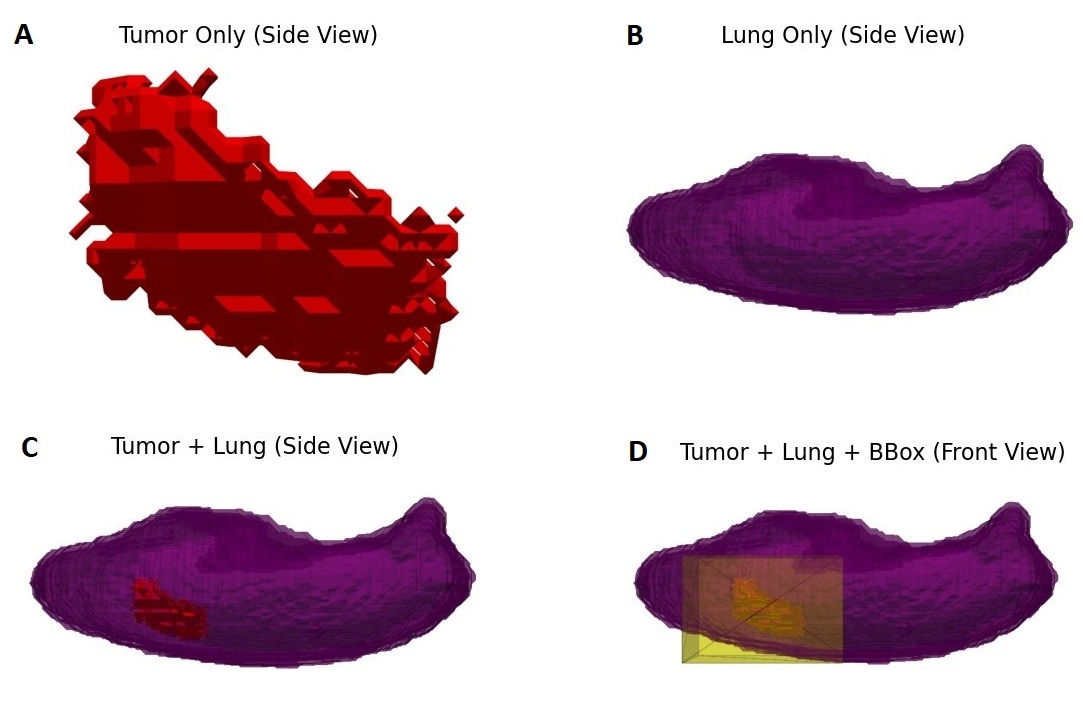}
\caption{Computational domain generated from CT scans. (A) Zoomed in MPM tumour. (B) Visceral pleura extracted from CT scan segmentation. 
(C) Tumour in the pleural space. For visualization purposes the parietal pleura is not shown. (D) Bounding box around the tumour, used to reduce the computational demand.} \label{fig2}
\end{figure}

The lung data are acquired from segmented CT scans using TotalSegmentator~\cite{wasserthal2023totalsegmentator,Armato2015} and used in the computational model, which is embedded in the agent-based CompuCell3D (see Fig.~\ref{fig2}B). In the pleural space, a cluster of epithelioid cancer cells is initialized to resemble an early stage of MPM (see Fig.~\ref{fig2}C).

\newpage
\subsection{Mathematical Models}
 For the simulation of tumour growth and cell proliferation we implement a CPM, as visualized in Fig.\ref{fig1}. The dynamics of oxygen ($C_O$), nutrients ($C_n$) and cytokines (IL6 and IL8) is described by a system of partial differential equations: 

\begin{equation}
\frac{dC_0}{dt} = D_O \nabla^2 C_O - S_0
\label{eq1}
\end{equation}

\begin{equation}
\frac{dC_n}{dt} = D_n \nabla^2 C_n - S_n
\label{eq2}
\end{equation}

\begin{equation}
\frac{dC_{\mathrm{IL6}}}{dt} = D_{\mathrm{IL6}} \nabla^2 C_{\mathrm{IL6}} 
 - \mu_{\mathrm{IL6}} \, C_{\mathrm{IL6}}
 \label{eq3}
\end{equation}

\begin{equation}
\frac{dC_{\mathrm{IL8}}}{dt} = D_{\mathrm{IL8}} \nabla^2 C_{\mathrm{IL8}}
 - \mu_{\mathrm{IL8}} \, C_{\mathrm{IL8}}
 \label{eq4}
\end{equation}

In these equations, $C_i$ represents concentrations, $D_i$ diffusion coefficients, S the consumption rate and $\mu$ the degradation rate for oxygen (O), nutrients (n) and the cytokines (IL6 and IL8).

\subsection{Bounding Box Mesh}
To model the proliferation of tumour cells in the pleural space, a regular 3D mesh was initialized for the spatial oxygen, nutrient and cytokine concentrations. Due to the large size of the lung and consequently the large simulation domain with around 9 million cells (CPM agents), it is unfeasible to perform multiple simulations and parametric variation analyses over the entire environment. To reduce computational complexity, we dynamically track the region of interest, namely the pleural space where the MPM tumour is growing, with an additional margin (applied to the computed bounding coordinates of all tumour cells) to account for diffusion. Dynamical tracking is based on the motility of the tumour and its expanding size, but occurs every 50 MCS to minimize overhead and keep the computational domain as small as possible, while capturing the relevant tumour interactions (see Fig.~\ref{fig2}D). 

\subsection{Implementation}
The sequential model is implemented in CompuCell3D, where each cell is represented as an individual agent, and uses the underlying Steppables functionality. Steppables are Python modules that execute specific functions at pre-defined simulation intervals (MCS). Examples of such Steppables that we use are \texttt{CellInitializationSteppable} (used to initialize the pleural space and tumour cells), \texttt{MitosisSteppable} (used to simulate mitosis) and \texttt{FiPySteppable}, the Steppable which performs dynamic domain partitioning and calls the parallel PDE solver. 

In the parallel PDE solver we use \texttt{FiPy} version 3.4.5 ~\cite{guyer2009fipy}, a finite-volume PDE solver parallelized via \texttt{mpi4py} version 4.0.3.~\cite{dalcin2021mpi4py} over \textit{n} cores, to solve the PDEs (eq.~\ref{eq1}--\ref{eq4}) governing the oxygen, nutrient and cytokine diffusion. The spatial domain within the dynamical bounding box is discretized by a structured 3D mesh. Time integration is handled using the first-order implicit Euler scheme and a linear iterative GMRES solver based on the Generalized Minimal Residual method ~\cite{saad1986gmres}. The resulting output is mapped on the pleural mask, representing the respective oxygen, nutrients, and cytokines concentration distributions. 

\subsection{Parallelization Strategy}
Due to the heavy computational load of the PDE solver, a separate script is developed outside of the sequential Steppables used by CompuCell3D, as can be seen in Fig.~\ref{fig1}. This script is called in parallel using \texttt{mpi4py} after the \texttt{FipySteppable} initializes the bounding box around the initial tumour in the pleural space and wrote the initial data (mesh, oxygen, nutrients, pleural mask and cytokines) to a shared file. 

Within the parallel solver, each process reads the shared file and takes an automatically partitioned domain using \texttt{FiPy’s} internal parallel structure. Once each process acquires a domain, it starts solving for that domain and sends back the updated concentrations upon completion. Once the PDEs solutions are obtained, the main process will verify the completion and distribute it to the CompuCell3D Steppables for computing cell-level processes such as proliferation, cell death, etc. This approach ensures that the PDE solver can scale to large 3D domains.

To evaluate the efficiency of the parallel implementation, we conducted a detailed performance analysis measuring the following key metrics:

\begin{enumerate}
\item The computational time required per timestep for both the parallel ($T_{Parallel}$) and serial ($T_{Serial}$) implementations;
 \item Parallel speedup $S_p$:
\begin{equation}
 S_p = \frac{T_{Serial}}{T_{Parallel}};
\end{equation}

\item Parallel efficiency $E_p$ on $p$ cores:
\begin{equation}
 E_p = \frac{S_p}{p};
\end{equation}

\item Load imbalance $f_{l,i}$, an important metric indicating how evenly the computational work is distributed among the cores:
\begin{equation}
f_{l,i} = \frac{t_{i}^{m} - \left(\frac{T_{i,s}}{p}\right)}{\langle t_{i}^{m} \rangle} = \frac{t_i^m}{\langle t_i \rangle} - 1 ,
\end{equation}
where $t_{i}^{m}$ is the maximum time for all p cores and $ t_i $ the average time it took for all p cores. %
%$\left(\frac{T_{i,s}}{P}\right)$ is the average time across all p cores, and $\langle t_{i}^{m} \rangle$ is the reference time (average time it took for 1 core).
\end{enumerate}

\section{Results}
In this section we present the performance analysis of the PDE solver under serial and parallel implementations. To compare the two implementations, we ran our simulation on three domain sizes ($100^3, 150^3$ and $200^3$) with the number of cores used ranging from 1 to 24.

\subsection{Computational Time }
\begin{figure}[!h]
\includegraphics[width=\textwidth]{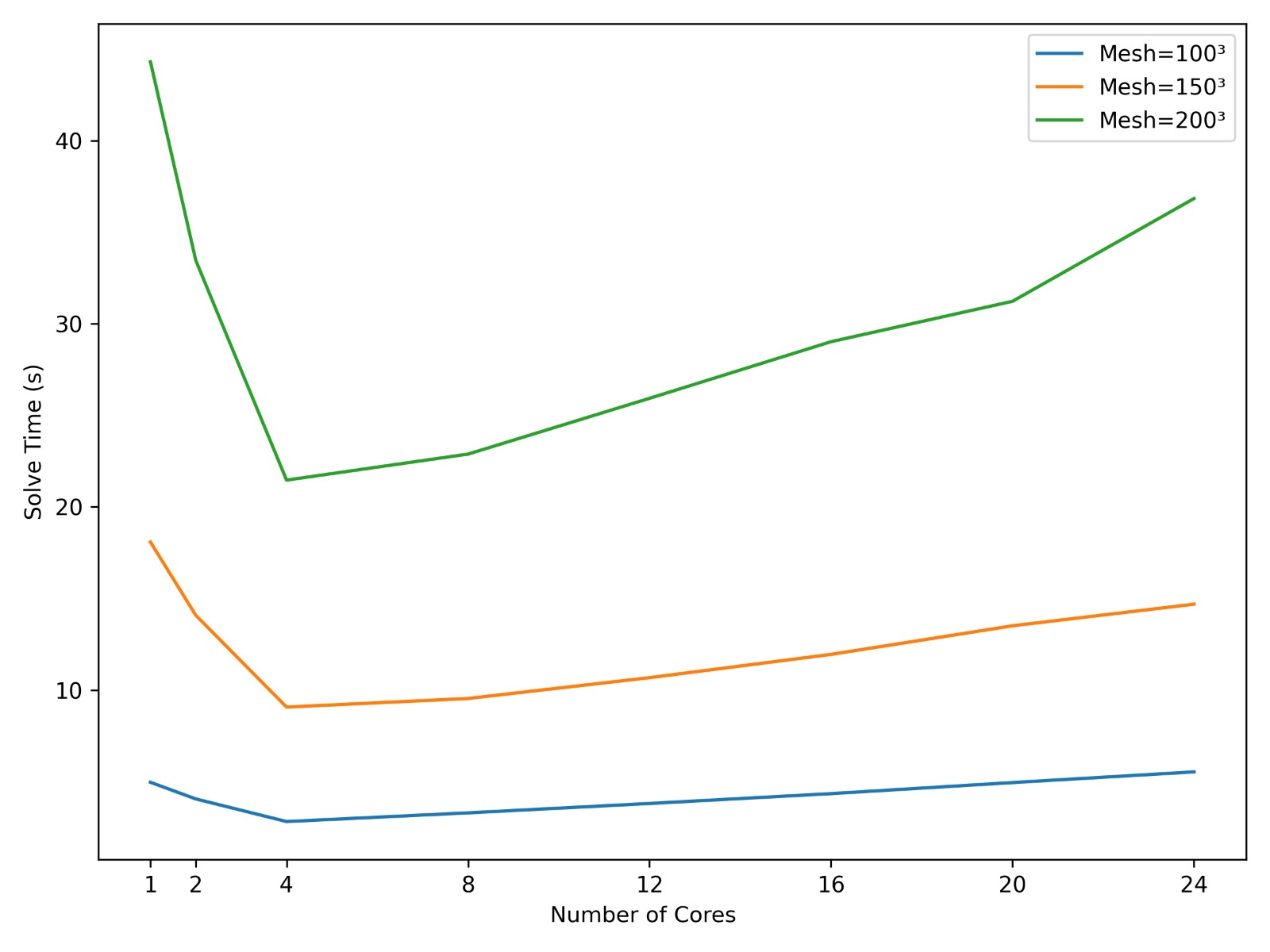}
\caption{PDE solver computational time per time step for three different domain sizes using different number of cores} \label{fig3}
\end{figure}

We start by evaluating our main metric, the computational time required per time step for both the parallel and serial implementations. In Fig.~\ref{fig3} we see how the computational time changes with increasing core counts for the three domain sizes. For all domain sizes, we notice a drastic change in computational time as we switch from a serial to parallel implementation, with the computational time decreasing as we increase the number of cores up to 4. 

This effect is stronger for larger domain sizes, with 4 cores proving to be the optimal number for these three domains. Beyond 8 cores, computational time is increasing, which can be explained by the growing communication overhead as the number of subdomains grows.

\subsection{Parallel Speedup}
\begin{figure}[!h]
\includegraphics[width=\textwidth]{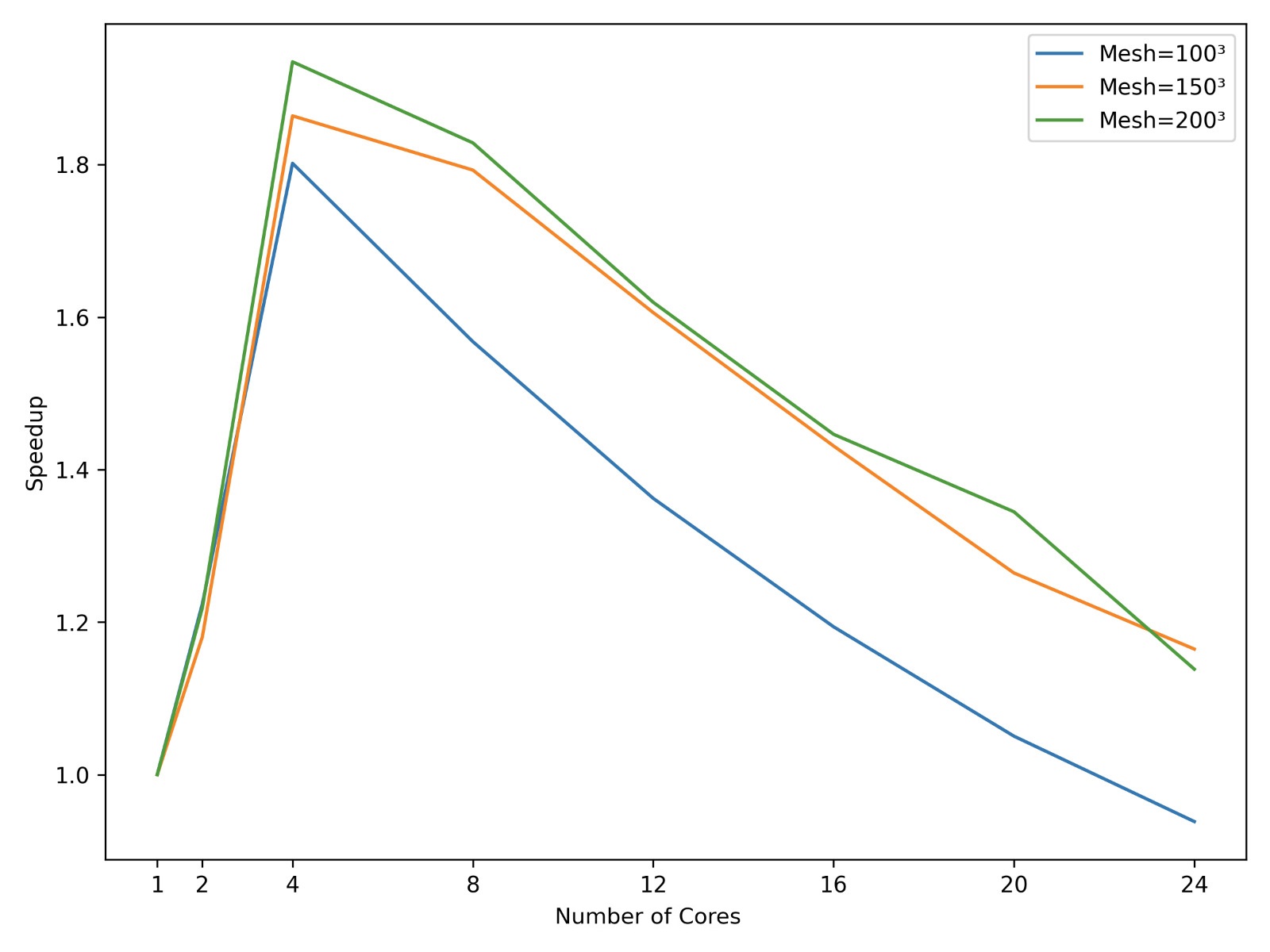}
\caption{Parallel speedup for various numbers of cores for the three respective domain sizes. } \label{fig4}
\end{figure}

In Fig.~\ref{fig4} we notice that for smaller number of cores, namely in the range of 2-4, the speedup increases sharply, reaching its peak for 4 cores with a speedup of 1.8 for the mesh of size $100^3$ and 1.95 for the mesh of size $200^3$. As the core count increases, the curves start to decline almost linearly. Despite this decline, only for the smallest domain at 24 cores the speedup is below 1, the larger domains maintain a slightly higher speedup due to having more compute work per core before the overhead becomes a dominant factor.

\subsection{Parallel Efficiency}
\begin{figure}[!h]
\includegraphics[width=\textwidth]{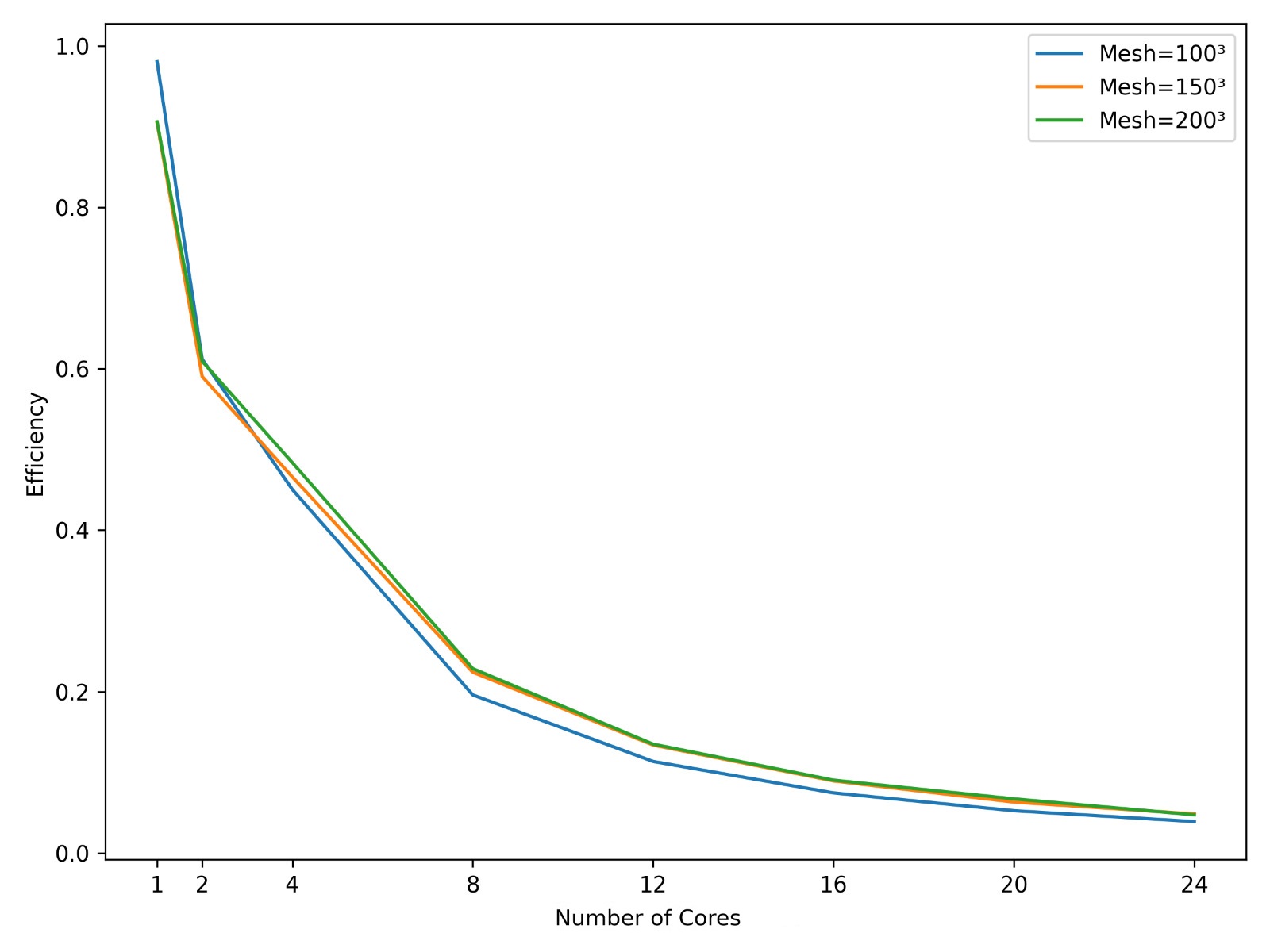}
\caption{Parallel efficiency for various numbers of cores for the three respective domain sizes.} \label{fig5}
\end{figure}

In Fig.~\ref{fig5} we see that for all domain sizes, parallel efficiency is close to 60\% on 2 cores, slightly deceasing to 50\% on 4 cores. With 8 or more cores, parallel efficiency is less than 20\%. This decline can be attributed to the ratio of communication time to computation time, which increases as more cores are added.
The difference between the three domain sizes is not significant, yet the larger domain size tends to maintain slightly higher efficiency compared to the smaller domains.

\subsection{Load Imbalance}
\begin{figure}[!h]
\includegraphics[width=\textwidth]{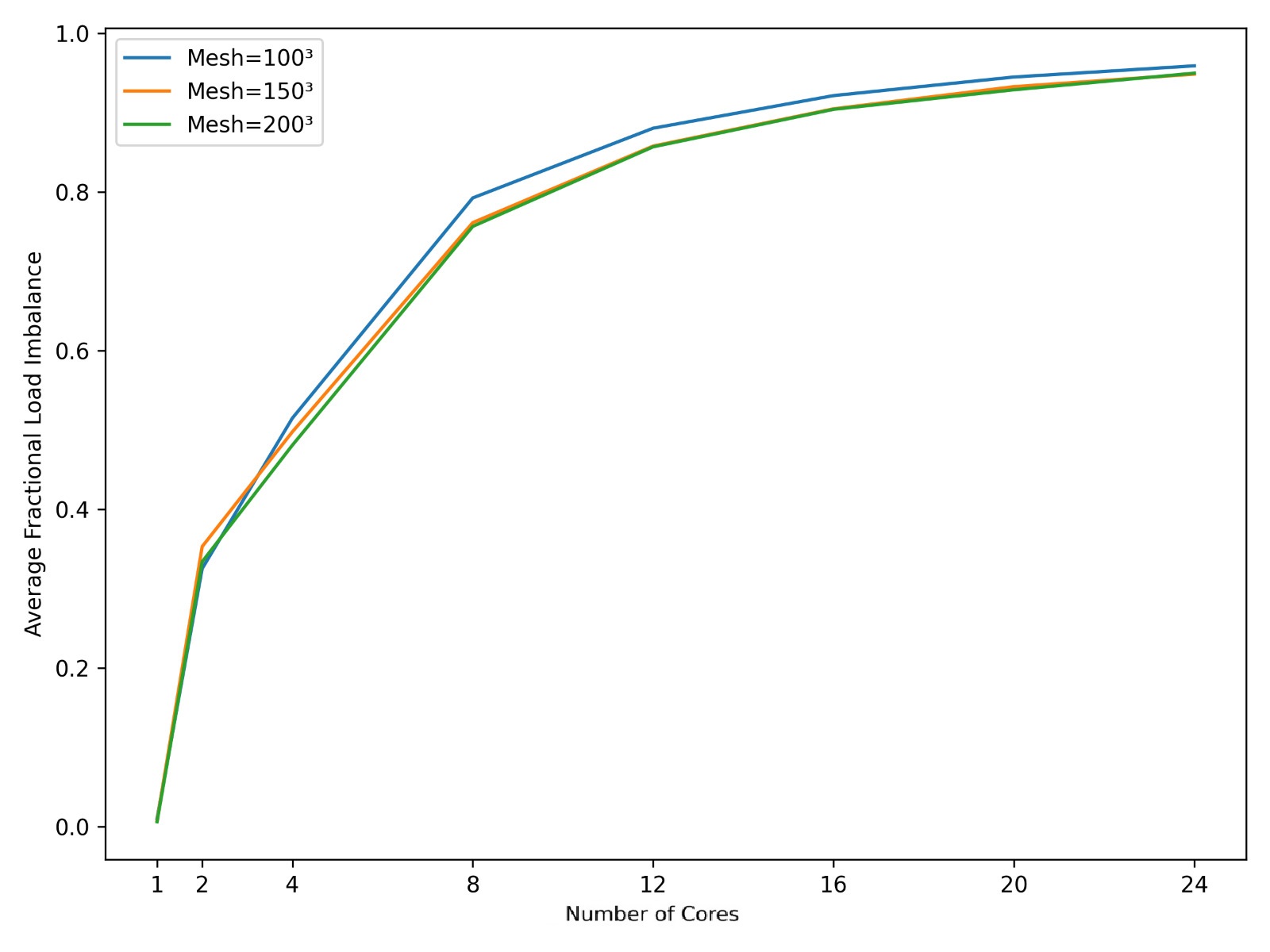}
\caption{Load imbalance for various numbers of cores for the three respective domain sizes. } \label{fig6}
\end{figure}

In Fig.~\ref{fig6} we notice that the simulations using fewer cores, namely 2 and 4, have a steep increase in the fractional load imbalance compared to the initial single-core simulation. This can be attributed to the fact that the assigned sub-domains per core vary in computational effort due to imperfect partitioning or varying boundary effects, resulting in some processes sitting idle whilst waiting for the other process to finish.

As the number of cores increases, the average fractional imbalance starts to approach 1, showing a more uneven distribution of workload among the cores. This waiting time, in turn, drives down the efficiency as can be seen in Fig.~\ref{fig5}.

\subsection{Impact of Dynamic Bounding Box}
\begin{figure}[!h]
\includegraphics[width=\textwidth]{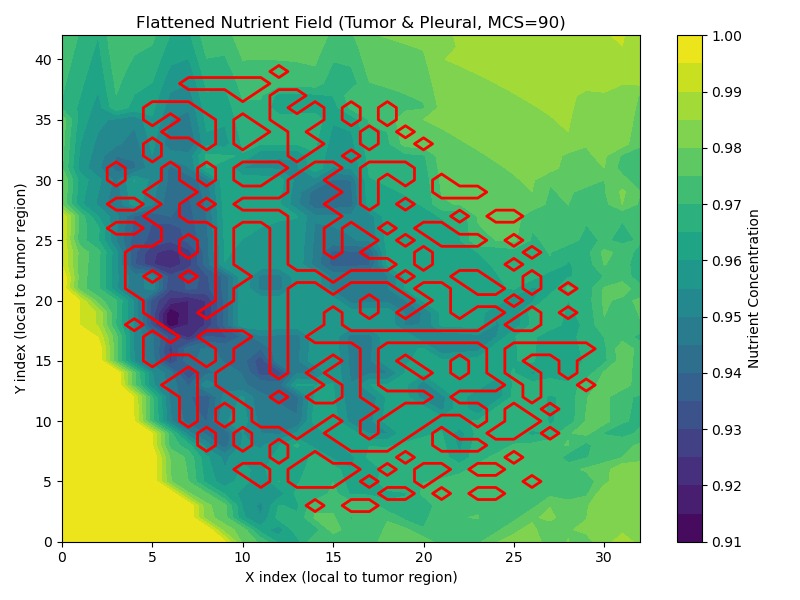}
\caption{A contour plot of the normalized nutrient concentrations in the pleural space (a 2D slice is shown), confined by the dynamically set bounding box. The perimeter lines of the tumour cells are highlighted in red.} \label{fig7}
\end{figure}

The dynamic bounding box plays a key role in reducing computational demand. By prioritizing the region of interest, it limits the number of cells that need to be processed. Solely due to the bounding box, the simulation can be run on a personal computer. Without it, the program cannot handle the computational demand of the whole pleural space and will result in severe performance issues or outright failure. This adaptive domain sizing decreases the memory overhead and reduces the number of cores required to run the simulation efficiently, thus reducing the overall computational time. Fig.~\ref{fig7} depicts the part of the pleural space inside the bounding box, for which the \texttt{FiPy} calculations are solved.

\section{Discussion}
Parallel computing can offer many advantages in the acceleration of complex simulations, whether it is working on a complex biofilm simulation or a predictive tumour growth model. Our results demonstrate that parallel implementation can significantly reduce computational time, yet it also comes with limitations. 

From our results we can conclude that switching from serial implementation to parallel implementation can reduce the computational time required for solving the PDEs. This general condition holds true for model conditions with cell counts in the range of 1 million to 8 million, with a linear speedup of 2 when using 4 cores.

At the same time, larger parallelization does not generalize to faster solving times as shown in our results (Fig.~\ref{fig4}). Use of multiple cores results in larger communication overhead and lower efficiency and speedup (Fig.~\ref{fig5} and~\ref{fig6}). The selection procedure for the number of cores to use depends on the domain size, and on the number of variables that need to be solved in that domain, as more variables result in more computational work.

The bounding box strategy has proved to be of key importance for this experiment, as it assisted in keeping the overall domain size manageable. It is important to note that due to its adaptive properties it greatly influences the domain size, and in a simulation where the tumour would grow aggressively and rapidly, the domain size could expand substantially, resulting in larger mesh sizes and consequently larger memory requirements. Thus, for future work it would be of interest to pair the bounding box mechanism with an automated core allocator based on domain size and solute components, to optimize solving speed and domain partitioning, thus acquiring optimal solving speed and memory usage throughout the whole simulation.

While multiple cores can reduce computation time, fine-grained load balancing, minimal interprocess communication, and efficient domain management are necessary to fully optimize parallel efficiency and avoid unwanted overhead.

\section{Conclusion}
We showed that the utilization of a parallel \texttt{FiPy} solver in a CPM based tumour growth model can accelerate the PDE computational time, with a speedup of 1.8-1.9 on 4 cores on mesh sizes $100^3$ - $200^3$. However, it is crucial that attention should be paid to communication overheads, mesh-level processes including domain partitioning and data exchange. The bounding box plays a key role in controlling the computational footprint of complex large-scale 3D computational domains by focusing the required computations on the region of interest within the pleural space. 

In the future, it is important to perform a scalability test for larger domains and a higher number of cores, improve load balancing, use adaptive meshing and automatically calculate the optimal number of cores before each execution of an MPI execute call. These improvements will greatly benefit the simulation of large-scale MPM tumours, by allowing more detailed and complex simulations at a lower computational cost. 
\bibliographystyle{splncs04}
\bibliography{citations}
\end{document}